%% file: main.tex
\documentclass[letterpaper, 10 pt, conference]{ieeeconf}

\IEEEoverridecommandlockouts 

\usepackage{tikz}
\usepackage{standalone}
\usepackage{bondgraphs}
\usepackage{amsmath}
\usepackage{amssymb}
\usepackage{mathtools}
\usepackage{url}
\usepackage{blox}
\usepackage{gensymb}
\usepackage{booktabs,amsfonts,dcolumn}
\usepackage{blindtext}
\usepackage{rotating}
\mathchardef\mhyphen="2D
\mathchardef\mslash="202F

\usepackage{textcomp}
\usepackage{hyperref}
\usepackage{multirow}
\usepackage{accents}
\usepackage{wrapfig}
\newcommand{\0}{\mathbf{0}}
\newcommand{\I}{\mathbf{I}}

\renewcommand{\top}{\mathsf{T}}
\newcommand{\R}{\mathbb{R}}

\DeclareRobustCommand{\ubar}[1]{\underaccent{\bar}{#1}}
\newcolumntype{L}[1]{>{\raggedright\arraybackslash}p{#1}}
\newcolumntype{C}[1]{>{\centering\arraybackslash}p{#1}}
\newcolumntype{R}[1]{>{\raggedleft\arraybackslash}p{#1}}
\usepackage{upgreek}
\usepackage[integrals]{wasysym}
\usepackage{mathdots}

\usepackage{tikz}
\usepackage{standalone}
\usepackage{bondgraphs}
\usetikzlibrary{shapes,arrows}
\usetikzlibrary{calc, patterns, decorations.pathmorphing, decorations.markings, shapes.geometric, arrows, positioning, angles}

\usepackage{amsthm}
\theoremstyle{definition}
\newtheorem{notation}{\bf Notation}
\newtheorem{problem}{\bf Problem}

\newtheorem{lemma}{\bf Lemma}

\newtheorem{definition}{\bf Definition}
\newtheorem{proposition}{\bf Proposition}

\DeclareSymbolFont{bfletters}{OT1}{cmr}{bx}{n}
\DeclareSymbolFontAlphabet{\mathbf}{bfletters}

\DeclarePairedDelimiter{\norm}{\lVert}{\rVert}
\DeclarePairedDelimiter{\abs}{\lvert}{\rvert}

\mathchardef\mhyphen="2D
\mathchardef\mslash="202F

\title{\LARGE \bf Barrier Pairs for Safety Control of Uncertain Output Feedback Systems}
\author{Binghan He and Takashi Tanaka
\thanks{This work was supported by the Air Force Office of Scientific Research under Grant \texttt{\small FA9550-20-1-0101}. \emph{(Corresponding author: Binghan He.)}}
\thanks{Binghan He was with the Oden Institute for Computational Engineering and Sciences, The University of Texas at Austin, Austin, TX 78712 USA. He is now with the Department of
Electrical Engineering and Computer Sciences, University of California, Berkeley, CA 94720 USA (e-mail: \texttt{\small binghan.he@berkeley.edu}).}
\thanks{Takashi Tanaka is with the Department of Aerospace Engineering and Engineering Mechanics, The University of Texas at Austin, Austin, TX 78712 USA (e-mail: \texttt{\small ttanaka@utexas.edu}).}}

\newcommand\copyrighttext{
\scriptsize 
Accepted for publication in American Control Conference (ACC)
\textcopyright 2023 AACC. Personal use of this material is permitted. Permission from AACC must be obtained for all other uses, in any current or future media, including reprinting/republishing this material for advertising or promotional purposes, creating new collective works, for resale or redistribution to servers or lists, or reuse of any copyrighted component of this work in other works.
DOI: \href{https://ieeexplore.ieee.org/document/10156174}{10.23919/ACC55779.2023.10156174}
}

\newcommand\copyrightnotice{
\begin{tikzpicture}[remember picture,overlay]
\node[anchor=south,yshift=10pt] at (current page.south)
{\fbox{\parbox{\dimexpr\textwidth-\fboxsep-\fboxrule\relax}{\copyrighttext}}};
\end{tikzpicture}
}

\begin{document}

\maketitle
\copyrightnotice
\vspace{-10pt}

\begin{abstract}
The barrier function method for safety control typically assumes the availability of full state information. Unfortunately, in many scenarios involving uncertain dynamical systems, full state information is often unavailable. In this paper, we aim to solve the safety control problem for an uncertain single-input single-output system with partial state information. First, we develop a synthesis method that simultaneously creates a barrier function and a dynamic output feedback safety controller. This safety controller guarantees that the unit sub-level set of the barrier function is an invariant set under the uncertain dynamics and disturbances of the system. Then, we build an identifier-based estimator that provides a state estimate affine to the uncertain model parameters of the system. To detect the potential risks of the system, a fault detector uses the state estimate to find an upper bound for the barrier function. The fault detector triggers the safety controller when the system's original action leads to a potential safety issue and resumes the original action when the potential safety issue is resolved by the safety controller.
\end{abstract}

\section{Introduction}

Barrier functions \cite{prajna2004safety, wieland2007constructive} are commonly used for safety control and verification of dynamical systems. However, the standard theory assumes that the full state information is available to compute the barrier function values. Unfortunately, full state information can only be estimated in many safety control scenarios involving disturbances and uncertainties, such as wearable robots \cite{thomas2021formulating} coupled with time-varying human dynamics and self-driving vehicles \cite{sadigh2018planning} in an uncertain environments.  

There are multiple methods \cite{tee2009barrier, ames2016control, thomas2018safety} for synthesizing a full state feedback controller that enforces a valid barrier function.
Thus, safety control for a system without full state measurements naturally begins with finding a state observer and then builds a safety controller using the state estimate from the state observer.
However, how to estimate the barrier function values using the state estimate from the state observer remains a question.
Although some observer-based output feedback controllers can enforce safety constraints without knowing the barrier function values \cite{feller2016robust, yu2019barrier}, we cannot blindly use these controllers at all times.
For example, a human operator may give a human assistive robot \cite{thomas2021formulating} an input that potentially violates some safety constraints of the human-robot coupled system. 
If we replace the original robot controller entirely with a safety controller, this robot will maintain safety but not assist the human operator.
For this reason, we must estimate the barrier function value in real-time so that the safety controller only corrects the original system when necessary.

In this paper, we aim to solve the safety control problem for an uncertain single-input single-output (SISO) system with partial state information. 
The main contributions of this paper are summarized as
follows.
\begin{itemize}
\item[(1)] 
In Sec.~\ref{sec:safety-control-0},
we develop a control synthesis method that simultaneously creates a barrier function and a dynamic output feedback safety controller. 
Using the controller parameter transformation scheme in \cite{scherer1997multiobjective}, our dynamic output feedback safety controller guarantees that the unit sub-level set of our barrier function is an invariant set with bounded model uncertainty and disturbance.
The proposed control synthesis method in this paper builds significantly upon the method in \cite{thomas2018safety}, which only focuses on full state feedback safety control.
\item[(2)] 
In Sec.~\ref{sec:fault-detection-0}, 
we propose a robust fault detector that consists of an identifier-based estimator \cite{morse1980global}.
Similar to the method in \cite{he2020robust}, this identifier-based estimator provides us with a robust state estimate, which helps us find the upper bound for our vector norm barrier function using partial state information.
However, unlike the estimator in \cite{he2020robust}, our fault detector in this paper does not require the system to be originally stable or have a stable static output feedback controller.
\item[(3)] 
In Sec.~\ref{sec:example}, we showcase our fault detector and safety controller, which work together to protect an uncertain SISO system from potential risks. 
In particular, we demonstrate how our fault detector helps us correctly trigger our safety controller when the system is about to violate the safety constraints and resume the original mission of the system when it is safe to do so.
\end{itemize}

\begin{notation}
We define 
$S_p \triangleq \begin{bmatrix} \I_{n \times n} & \0_{n \times n} \end{bmatrix} $
and
$S_k \triangleq \begin{bmatrix} \0_{n \times n} & \I_{n \times n} \end{bmatrix} $
as two selection matrices, which extract the plant state $x_p$ and the controller state $x_k$ from the closed-loop state vector $x_\mathsf{CL}$. 
Supposing $P \succ \0$,
\begin{equation}
\norm{\star}_{P} \triangleq \sqrt{\star ^ \top P \star}
\end{equation}
is a vector norm function based on $P$. In our LMIs, we define $\mathsf{He} \{ \star \} \triangleq \star + \star ^ \top$.
\end{notation}

\section{Preliminaries}

In this section, we introduce models of an uncertain dynamical system $\Sigma_p$ and a full-order dynamic output feedback controller $\Sigma_k$. The uncertain dynamical system is described as a polytopic linear differential inclusion (PLDI) \cite{boyd1994linear}. Then we give an overview of barrier pairs, which will be used for safety verification and control of the uncertain dynamical system. Based on the concept of barrier pairs, we present our formal problem statement.

\subsection{State Model}

In this paper, we consider a SISO system. We will explore how we can extend our study to a multiple-input multiple-output (MIMO) system in the future. 

First, assume that the transfer function of our system $\Sigma_p$ from input $u$ to output $y$ is expressed as 
\begin{equation} \label{eq:tf}
G_p (s) = \frac{y (s)}{u (s)} = \frac{\hfill \beta_{1} s ^ {n-1} + \cdots + \beta_{n-1} s + \beta_{n}}{s ^ n + \alpha_{1} s ^ {n-1} + \cdots + \alpha_{n-1} s + \alpha_{n}}, 
\end{equation}
where $\alpha_1, \, \alpha_2, \, \cdots, \, \alpha_n$ and $\beta_1, \, \beta_2, \, \cdots, \, \beta_n$ are the uncertain parameters of our plant.
Let us define
\begin{equation}
A_0 \triangleq \begin{bmatrix*}[l] \0_{(n-1) \times 1} ^ \top & 0 \\ \I_{(n - 1) \times (n - 1)} & \0_{(n-1) \times 1} \end{bmatrix*} \ 
\text{and} \ 
c_0 \triangleq \begin{bmatrix*}[l] \0_{1 \times (n-1)} & 1 \end{bmatrix*}. 
\end{equation}
In Sec.~\ref{sec:identifier}, we will conduct a robust fault detection of $\Sigma_p$ using an identifier-based estimator \cite{morse1980global}. 
In order to build this identifier-based estimator, we need to use the following state space realization of $\Sigma_p$: 
\begin{align}
\dot{x}_{p}
& 
= 
A_0 x_{p} 
+ 
\begin{bmatrix}
b_y & b_u
\end{bmatrix}
\begin{bmatrix}
y \\ u
\end{bmatrix}
\label{eq:model-siso} \\
y 
& 
= 
c_0 x_{p} 
+ 
w \notag 
\end{align}
where $b_y \triangleq [ - \alpha_n \ \cdots \ - \alpha_{2} \ - \alpha_1 ] ^ \top$ and $b_u \triangleq [ \beta_n \ \cdots \ \beta_{2} \ \beta_1 ] ^ \top$ contain all the uncertain plant parameters and $w$ is a disturbance signal.
Let us suppose that
\begin{equation} \label{eq:c-limit}
\begin{aligned}
b_y & = \bar{b}_y + \sum_{i} ^ {n_p} \delta_i \cdot \theta_i ^ y \cdot \tilde{b}_i \\
b_u & = \bar{b}_u + \sum_{i} ^ {n_p} \delta_i \cdot \theta_i ^ u \cdot \tilde{b}_i 
\end{aligned}
\end{equation}
where 
$\bar{b}_y \triangleq [ - \bar{\alpha}_n \ \cdots \ - \bar{\alpha}_{2} \ - \bar{\alpha}_1 ] ^ \top$ and $\bar{b}_u \triangleq [ \bar{\beta}_n \ \cdots \ \bar{\beta}_{2} \ \bar{\beta}_1 ] ^ \top$ are the nominal plant parameters,
$n_p$ is the number of dimensions of the parameter uncertainties,
$\tilde{b}_i \in \R ^ {n \times 1}$, $\theta_i ^ y, \, \theta_i ^ u \in \R$ describe the direction of the parameter uncertainties in each dimension, 
$\delta_i \in \R$ is a scalar variable such that $\abs{\delta_i} \leq 1$ for all $i = 1, \, 2, \, \cdots, \, n_p$.
Based on \eqref{eq:c-limit}, $\begin{bmatrix} b_y & b_u \end{bmatrix}$ is in a polytopic region in $\R ^ {n \times 2}$ and \eqref{eq:model-siso} becomes a PLDI.

Next, let the state-space model of a full-order dynamic output feedback controller $\Sigma_k$ be
\begin{alignat}{1}
\dot{x}_{k} & = A_k x_{k} + b_k y \label{eq:controller-siso} \\
          u & = c_k x_{k} \notag 
\end{alignat}
where $A_k \in \R ^ {n \times n}$, $b_k \in \R ^ {n \times 1}$, and $c_k \in \R ^ {1 \times n}$ are the controller parameters to be determined.

Let us define 
\begin{alignat}{4}
&
A_\mathsf{CL} 
&&
\triangleq
\begin{bmatrix*}
\hat{A}_p & \bar{b}_u c_k \\
b_k c_0 & A_k
\end{bmatrix*}
, 
\quad
&& B_{w p} 
&&
\triangleq
\begin{bmatrix}
\bar{b}_y & \tilde{b}_p \\ b_k & \0
\end{bmatrix}
,
\label{state-OL}
\\
&
C_q 
&&
\triangleq 
\begin{bmatrix}
\Theta_y ^ \top c_0 & \Theta_u ^ \top c_k
\end{bmatrix}
, 
\quad
&& D_{w p} 
&&
\triangleq
\begin{bmatrix*} \Theta_y ^ \top & \0 \end{bmatrix*}
, 
\notag
\end{alignat}
where $\hat{A}_p \triangleq A_0 + \bar{b}_y c_0$, $\tilde{b}_p \triangleq [ \tilde{b}_1 \ \tilde{b}_2 \ \cdots \ \tilde{b}_{n_p} ]$, $\Theta_y \triangleq [ \theta_1 ^ y \ \theta_2 ^ y \ \cdots \ \theta_{n_p} ^ y ]$ and $\Theta_u \triangleq [ \theta_1 ^ u \ \theta_2 ^ u \ \cdots \ \theta_{n_p} ^ u ]$.
Combining \eqref{eq:model-siso} and \eqref{eq:controller-siso}, our closed-loop system $\Sigma_\mathsf{CL}$ for $u = \Sigma_k (y)$ is described as
\begin{alignat}{6} \label{eq:model-CL}
&
\dot{x}_\mathsf{CL} 
&& 
= 
A_\mathsf{CL}  
&&
x_\mathsf{CL} 
&& 
+
B_{w p} 
&&
\begin{bmatrix} w \\ p \end{bmatrix}  \\
&
q              
&& 
= 
C_q      
&&
x_\mathsf{CL} 
&& 
+
D_{w p} 
&&
\begin{bmatrix} w \\ p \end{bmatrix}
\notag
\end{alignat}
where 
$x_\mathsf{CL} \triangleq [ x_{p} ^ \top \ x_{k} ^ \top ] ^ \top$, $p = [p_1 \ p_2 \ \cdots \ p_{n_p}] ^ \top$, $q = [q_1 \ q_2 \ \cdots \ q_{n_p}] ^ \top$ and $p_i = \delta_i q_i$ for all $i = 1, \, 2, \, \cdots, \, n_p$. 

\subsection{Barrier Pairs}

The concept of barrier pairs \cite{thomas2018safety} describes the relationship between a barrier function and a feedback controller in a safety control problem. 
In this paper, we extend the definition of barrier pairs to output feedback systems.
\begin{definition} \label{def:bp}
A barrier pair is a pair $(\mathbf{B},\ \Sigma_k)$ consisting of a barrier function $\mathbf{B}: x_\mathsf{CL} \rightarrow \R$ and a controller $\Sigma_k$ satisfying the following conditions:
\begin{itemize}
\item[(a)] $\varepsilon \leq \mathbf{B} (x_\mathsf{CL}) \leq 1, \, u = \Sigma_k (y) \implies \dot{\mathbf{B}} (x_\mathsf{CL}) < 0$,
\vspace{3pt}
\item[(b)] $\mathbf{B} (x_\mathsf{CL}) \leq 1 \implies x_p \in \mathcal{X}_s, \ \Sigma_k (y) \in \mathcal{U}$.
\vspace{1pt}
\end{itemize}
In particular, $x_p \in \mathcal{X}_s$ and $u \in \mathcal{U}$ are the state and input constraints.
Intuitively, (a) and (b) mean the invariance and constraint satisfaction properties of a barrier pair.
\end{definition}

\begin{figure}
\centering
\def\svgwidth{0.49\textwidth}
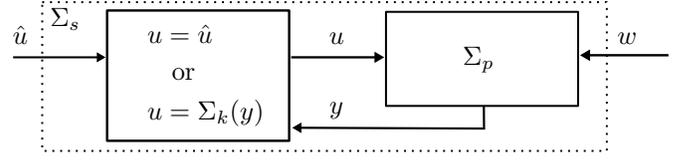
\caption{In this paper, our switching system $\Sigma_s$ chooses either an original input $u = \hat{u}$ of $\Sigma_p$ or a known-to-be-safe input $u = \Sigma_k (y)$ based on a barrier function $B$.}
\label{fig:problem}
\end{figure}

\subsection{Problem Statement}

In this paper, we consider an uncertain dynamical system $\Sigma_p$ described in \eqref{eq:model-siso} and \eqref{eq:c-limit} with assumption $\abs{w} \leq \bar{w}$ and safety constraints $x_p \in \mathcal{X}_s$ and $u \in \mathcal{U}$. 
\begin{problem}
Assuming that $y$ and $u$ are the only available measurements of $\Sigma_p$ and $x_p = \0$ at $t = 0$, find a barrier pair $(B, \, \Sigma_k)$ and a switching system $\Sigma_s$ where
$(B, \, \Sigma_k)$ satisfies conditions (a) and (b) in Definition~1 for $\Sigma_p$ and where $\Sigma_s$ switches $\Sigma_p$ from its original input $u = \hat{u}$ to $u = \Sigma_k (y)$ whenever $B$ is close to $1$ and switches back to $u = \hat{u}$.
\end{problem}

Fig.~\ref{fig:problem} shows a block diagram of $\Sigma_s$ that is similar to the switching systems proposed in \cite{wieland2007constructive, thomas2018safety}. This type of switching system seeks to maintain the safety of $\Sigma_p$ through a minimum intervention in its original input $u = \hat{u}$. However, unlike the previous works, the full state of $\Sigma_p$ is not available in our problem. Therefore, our switching system cannot switch between $u = \hat{u}$ and $u = \Sigma_k (y)$ based on the exact value of our barrier function $\mathbf{B}$. 

To address this limitation, we will solve our problem in two steps. In Sec.~\ref{sec:safety-control-0}, we show a synthesis method that creates a barrier pair for $\Sigma_p$. In Sec.~\ref{sec:fault-detection-0}, we propose a robust fault detector that finds an upper bound for $\mathbf{B}$ using only the measurements of $y$ and $u$. Based on this upper bound for $\mathbf{B}$, we will construct $\Sigma_s$ in Sec.~\ref{sec:switching}.

\section{Barrier Pair Synthesis} \label{sec:safety-control-0}

In this section, we focus on the barrier pair synthesis for our system $\Sigma_p$ described in \eqref{eq:model-siso} and \eqref{eq:c-limit}. First we formulate conditions (a) and (b) in Definition~\ref{def:bp} as linear matrix inequality (LMI) constraints. Then we introduce our LMI optimization problems for barrier pair synthesis.

\subsection{LMIs for Invariance Property}

In this paper, we define our barrier function as a vector norm function
\begin{equation} \label{eq:barrier-function}
\mathbf{B} (x_\mathsf{CL}) \triangleq \norm{x_\mathsf{CL}}_{P}.
\end{equation}
Let us partition $P$ and $Q \triangleq P ^ {- 1}$ as
\begin{equation} \label{eq:P}
P = 
\begin{bmatrix*}[l] 
X & V \\ 
V ^ \top & \star 
\end{bmatrix*}
\quad \text{and} \quad
Q =
\begin{bmatrix*}[l] 
Y & W \\ 
W ^ \top & \star 
\end{bmatrix*}
,
\end{equation}
where $X, \, Y, \, V, \, W \in \R ^ {n \times n}$. In addition, we define
\begin{equation}
\Pi_1 \triangleq 
\begin{bmatrix*}[l]
\I & X \\
\0 & V ^ \top
\end{bmatrix*}
\quad \text{and}
\quad
\Pi_2 \triangleq 
\begin{bmatrix*}[l]
Y        & \I \\
W ^ \top & \0
\end{bmatrix*}
,
\end{equation}
where $\Pi_1 = P \Pi_2$ and $\Pi_2 = Q \Pi_1$. In Proposition~\ref{prop:CL-convex}, we will use $\Pi_1$ and $\Pi_2$ to perform a controller parameter transformation \cite{scherer1997multiobjective} and derive the LMI constraints for condition (a) in Definition~\ref{def:bp}.
\begin{proposition} \label{prop:CL-convex}
Supposing that $\abs{w} \leq \bar{w}$, there exist a barrier function  
$\mathbf{B} (x_\mathsf{CL}) = \norm{x_\mathsf{CL}}_{P}$ and a controller $\Sigma_k$ in the form of \eqref{eq:controller-siso} such that $( \mathbf{B} (x_\mathsf{CL}), \, \Sigma_k)$ satisfies condition~(a) in Definition~\ref{def:bp}, if there exist $X \succ \0$, $Y \succ \0$, $E \in \R ^ {n \times n}$, $F \in \R ^ {n \times 1}$, $G \in \R ^ {1 \times n}$, and $\mu_w, \, \mu_1, \, \mu_2, \, \cdots, \, \mu_{n_p} \geq 0$ such that 
\begin{equation} \label{eq:z1-z2}
\begin{bmatrix} 
 Y & \I \\ 
\I &  X 
\end{bmatrix} 
\succ 0. 
\end{equation}
and
\begin{equation} 
\begin{bmatrix*}
H_A & \star     & \star \\
H_B ^ \top &   - \mathbf{M}_{w p}     & \star \\
\mathbf{M}_p H_C &   \mathbf{M}_p D_{w p} & - \mathbf{M}_p
\end{bmatrix*}
\prec 0, 
\label{eq:LMI-lyapunov}
\end{equation}
where $\mathbf{M}_p \triangleq \mathsf{diag} ( \mu_1, \ \mu_2, \ \cdots, \ \mu_{n_p} )$, $\mathbf{M}_{w p} \triangleq \mathsf{diag} ( \mu_w, \ \mathbf{M}_p )$,
\begin{flalign}
H_A & \triangleq
\text{\small $
\mathsf{He} \bigg\{
\begin{bmatrix*}
\hat{A}_p Y + \bar{b}_u G & \hat{A}_p \\
  E & X \hat{A}_p + F c_0
\end{bmatrix*}
\bigg\}
+ 
\mu_w \frac{\bar{w} ^ 2}{\varepsilon ^ 2} 
\begin{bmatrix*}
 Y & \I \\ 
\I &  X
\end{bmatrix*}
$}
,
& \\
H_B & \triangleq
\text{\small $
\begin{bmatrix*}[r]
\bar{b}_y & \tilde{b}_p \\
F + X \bar{b}_y & X \tilde{b}_p 
\end{bmatrix*}
$}
, \quad
H_C   \triangleq
\text{\small $
\begin{bmatrix*}[r]
\Theta_u ^ \top G + \Theta_y ^ \top c_0 Y & \Theta_y ^ \top c_0
\end{bmatrix*}
$}
,
& \notag \\
\text{and} && \notag \\
E & \triangleq V A_k W ^ \top + F c_0 Y + X \bar{b}_u G + X \hat{A}_p Y, & \label{eq:variable-1} \\
F & \triangleq V b_k, \qquad G \triangleq c_k W ^ \top. & \notag
\end{flalign}
\end{proposition}

\begin{proof}
Notice that \eqref{eq:LMI-lyapunov} is obtained by performing a congruence transformation with $\mathsf{diag} ( \Pi_1, \ \I, \ \I ) $ on
\begin{equation} \label{eq:LMI-lyapunov-0}
\begin{bmatrix*}
\Phi_{Q} + \mu_w \frac{\bar{w} ^ 2}{\varepsilon ^ 2} Q & \star & \star \\
B_{wp} ^ \top & - \mathbf{M}_{w p}  & \star \\
\mathbf{M}_p C_q Q & \mathbf{M}_p D_{w p} & - \mathbf{M}_p \\
\end{bmatrix*}
\prec \0,
\end{equation}
where $\Phi_{Q} \triangleq A_\mathsf{CL} Q + Q A_\mathsf{CL} ^ \top$.
Thus, we focus on the following two steps to complete the proof.
First, we will show that condition (a) in Definition~\ref{def:bp} holds for $( \mathbf{B} (x_\mathsf{CL}), \, \Sigma_k)$ if \eqref{eq:LMI-lyapunov-0} holds. Then, we will show that \eqref{eq:LMI-lyapunov-0} holds if \eqref{eq:z1-z2} and \eqref{eq:LMI-lyapunov} hold.

In the first step, we consider $\mathbf{B} ^ 2 (x_\mathsf{CL}) = x_\mathsf{CL} ^ \top P x_\mathsf{CL}$ as a quadratic Lyapunov function candidate for our closed-loop system with $u = \Sigma_k (y)$. Then, $( \mathbf{B} (x_\mathsf{CL}), \, \Sigma_k)$ satisfies condition~(a) in Definition~\ref{def:bp} for $\Sigma_p$ in \eqref{eq:model-siso} and \eqref{eq:c-limit} under the assumption $\abs{w} \leq \bar{w}$
if and only if $\frac{\mathsf{d} \, B ^ 2 (x_\mathsf{CL})}{\mathsf{d} \, t} < 0$, or equivalently
\begin{equation} \, \label{eq:B-dot}
\begin{bmatrix}
x_\mathsf{CL} \\
w \\
p
\end{bmatrix}
^ \top 
\begin{bmatrix*}
A_\mathsf{CL} ^ \top P + P A_\mathsf{CL} & \star & \star \\
B_{w} ^ \top P & \0 & \star \\
B_{p} ^ \top P & \0 & \0 \\
\end{bmatrix*}
\begin{bmatrix}
x_\mathsf{CL} \\
w \\
p
\end{bmatrix}
< 0,
\end{equation}
for all $x_\mathsf{CL}$, $w$, and $p$ that
\begin{equation} \label{eq:s-condition}
\begin{aligned}
x_\mathsf{CL} ^ \top P x_\mathsf{CL} \geq \varepsilon ^ 2, \quad w ^ 2 \leq \bar{w} ^ 2, \quad \text{and} \quad p_i ^ 2 \leq q_i ^ 2, \\
\forall \ i = 1, \, \cdots, \, n_p.
\end{aligned} 
\end{equation}
Using the S-procedure, \eqref{eq:B-dot} holds under the conditions in \eqref{eq:s-condition} if there exist $\mu_\mathsf{CL}, \, \mu_w, \, \mu_1, \, \cdots, \, \mu_{n_p} \geq 0$ such that for all $x_\mathsf{CL}$, $w$, and $p$ that
\begin{equation} \label{eq:LMI-lyapunov-1}
\begin{bmatrix}
x_\mathsf{CL} \\
w \\
p
\end{bmatrix}
^ \top 
H_P
\begin{bmatrix}
x_\mathsf{CL} \\
w \\
p
\end{bmatrix}
+ \mu_w \bar{w} ^ 2 - \mu_\mathsf{CL}
< 0,
\end{equation}
where 
\begin{equation} \label{eq:LMI-lyapunov-2}
H_P \triangleq
\begin{bmatrix*}
\Phi_{P} + \frac{\mu_\mathsf{CL}}{\varepsilon ^ 2} P + C_q ^ \top \mathbf{M}_p C_q & \star \\
B_{w p} ^ \top P & - \mathbf{M}_{w p} + D_{w p} ^ \top \mathbf{M}_p D_{w p} 
\end{bmatrix*}
\end{equation}
and $\Phi_{P} \triangleq A_\mathsf{CL} ^ \top P + P A_\mathsf{CL}$.
Then, \eqref{eq:LMI-lyapunov-1} holds if $H_P \prec \0$ and $\mu_\mathsf{CL} = \mu_w \bar{w} ^ 2$.
Through a congruence transformation with $\mathsf{diag} ( Q, \ \I )$ on $H_P$ and the Schur complement, $H_P \prec \0$ for $\mu_\mathsf{CL} = \mu_w \bar{w} ^ 2$ is equivalent to \eqref{eq:LMI-lyapunov-0}.

Now, let us establish the second step. 
Since \eqref{eq:LMI-lyapunov} is obtained by performing a congruence transformation with $\mathsf{diag} ( \Pi_1, \ \I, \ \I ) $ on \eqref{eq:LMI-lyapunov-0}, \eqref{eq:LMI-lyapunov-0} holds if there exists non-singular $\mathsf{diag} ( \Pi_1, \ \I, \ \I ) $ such that \eqref{eq:LMI-lyapunov} holds.
Obviously, $\mathsf{diag} ( \Pi_1, \ \I, \ \I ) $ is non-singular if and only if $V$ is non-singular. Since $P Q = \I$, we have $V W ^ \top = \I - X Y$. Then, there exists non-singular $V$ and $W$ if $\I - X Y$ is non-singular.
$\I - X Y$ is non-singular if $Y \succ \0$ and $X - Y ^ {- 1} \succ \0$. Using the Schur complement, $Y \succ \0$ and $X - Y ^ {- 1} \succ \0$ if \eqref{eq:z1-z2} holds.
Consequently, we obtain that there exists non-singular $\mathsf{diag} ( \Pi_1, \ \I, \ \I ) $ if \eqref{eq:z1-z2} holds.

Therefore, condition (a) in Definition~\ref{def:bp} holds for $( \mathbf{B} (x_\mathsf{CL}), \, \Sigma_k)$ if there exist $X \succ \0$, $Y \succ \0$, $E \in \R ^ {n \times n}$, $F \in \R ^ {n \times 1}$, $G \in \R ^ {1 \times n}$, and $\mu_w, \, \mu_1, \, \mu_2, \, \cdots, \, \mu_{n_p} \geq 0$ such that \eqref{eq:z1-z2} and \eqref{eq:LMI-lyapunov} hold.
\end{proof}

If we define the scalar variables $\mu_w, \, \mu_1, \, \mu_2, \, \cdots, \, \mu_{n_p}$ a priori, \eqref{eq:LMI-lyapunov} becomes an LMI in $( X, \ Y, \ E, \ F, \ G )$.
Even though \eqref{eq:LMI-lyapunov} is not an LMI with respect to these variables jointly, 
we can consider methods such as D-K iteration \cite{balas1993mu} for searching the values of these scalar variables.

\subsection{LMIs for State and Input Limits}

Now, let us focus on condition (b) in Definition~\ref{def:bp} for barrier function $\mathbf{B} (x_\mathsf{CL}) = \norm{x_\mathsf{CL}}_{P}$ and safety controller $\Sigma_k$ in the form of \eqref{eq:controller-siso}. In this paper, let us define $\mathcal{X}_s$ and $\mathcal{U}$ as
\begin{alignat}{4}
\mathcal{X}_s & \triangleq \{x_p \ & : \ && |f_i ^ \top x_p| & \leq 1, \ i = 1, \, 2, \, \cdots, \, n_f \}, \\
\mathcal{U}   & \triangleq \{  u \ & : \ &&           |u| & \leq \bar{u} \}. 
\end{alignat}
Since $f_i ^ \top x_p = f_i ^ \top S_p x_\mathsf{CL}$ and $u = c_k S_k x_\mathsf{CL}$, condition~(b) in Definition~\ref{def:bp} holds for $( \mathbf{B} (x_\mathsf{CL}), \, \Sigma_k)$ if
\begin{alignat}{4}
& f_i ^ \top && S_p Q S_p ^ \top && f_i && \leq 1, \quad \text{for} \ i = 1, \, 2, \, \cdots, \, n_f, \label{eq:LMI-xb} \\
& c_k && S_k Q S_k ^ \top && c_k ^ \top && \leq \bar{u} ^ 2. \label{eq:LMI-ub-0}
\end{alignat}
Since $\Sigma_k$ is undetermined (i.e., $A_k$, $b_k$ and $c_k$ are also decision variables), \eqref{eq:LMI-ub-0} is not an LMI. 
We address this issue as follows.
Since $Q = P ^ {- 1}$, we can use the Schur complement to obtain that \eqref{eq:LMI-ub-0} holds if 
\begin{equation} \label{eq:LMI-ub-1}
\begin{bmatrix}P & \star \\c_k S_k & \bar{u} ^ 2\end{bmatrix}\succeq \0.
\end{equation}
By performing a congruence transformation with $\mathsf{diag} ( \Pi_2, \ \I) $ on \eqref{eq:LMI-ub-1}, we obtain that \eqref{eq:LMI-ub-0} holds if
\begin{align}
&
\begin{bmatrix}
Y  & \star & \star \\
\I &     X & \star \\
G ^ \top  & \0 & \bar{u} ^ 2
\end{bmatrix}
\succeq \0, \label{eq:LMI-ub}
\end{align}
where \eqref{eq:LMI-ub} is an LMI in our new variable set $( X, \ Y, \ E, \ F, \ G )$ introduced in Proposition~\ref{prop:CL-convex}. 
In addition, we can use the Schur complement to obtain that \eqref{eq:LMI-ub} also implies condition \eqref{eq:z1-z2} in Proposition~\ref{prop:CL-convex}.

\subsection{Barrier Pair Construction}

Through a convex optimization
\begin{equation} \label{eq:optimization-volume}
\begin{aligned}
& \underset{X, \, Y, \, E, \, F, G}{\mathsf{maximize}}
& & \mathsf{log}(\mathsf{det} (Y)) \\
& \mathsf{subject \ to} & & X \succ 0, \, Y \succ 0, \\
&&& \eqref{eq:LMI-lyapunov}, \, \eqref{eq:LMI-xb}, \, \eqref{eq:LMI-ub}, \\
\end{aligned}
\end{equation}
we obtain a solution of $( X, \ Y, \ E, \ F, \ G )$ that maximize the volume of the $x_p$ space projection $\{ x_p : x_p = S_p x_\mathsf{CL}, \ \mathbf{B} (x_\mathsf{CL}) \leq 1 \}$ of the unit sub-level set of $\mathbf{B} (x_\mathsf{CL})$. 

There are multiple methods to construct a controller $\Sigma_k$ based on a solution of $( X, \ Y, \ E, \ F, \ G )$.
For example, Ref.~\cite[Lemma~7.9]{dullerud2013course} defines $V V ^ \top = X - Y ^ {- 1}$ and $W = - Y V$. 
After obtaining $V$ and $W$, we can construct our controller parameters $A_k$, $b_k$, and $c_k$ according to \eqref{eq:variable-1}.

\section{Robust Fault Detection} \label{sec:fault-detection-0}

Although barrier pair $(\mathbf{B}, \, \Sigma_k)$ is obtained in the previous section, the calculation of $\mathbf{B}$ relies on knowing the full state $x_\mathsf{CL}$ of the closed-loop system. Since $x_p$ is not available, the true value of $\mathbf{B}$ is unknown. In this section, we will introduce a robust fault detector that provides an upper bound for $\mathbf{B}$ using only the measurements of $y$ and $u$.

\subsection{Identifier-Based Estimator} \label{sec:identifier}

In \cite{morse1980global}, the concept of an identifier-based estimator was developed for the purpose of model identification and adaptive control. However, as a byproduct, it also provides us with a robust state estimate $\hat{x}_p$ to the model uncertainty of $\Sigma_p$. The identifier-based estimator for our system $\Sigma_p$ is a pair of sensitivity function filters
\begin{equation} \label{eq:model-0}
\begin{aligned}
\dot{z}_y & = A_{z} ^ \top z_y + c_0 ^ \top y \\
\dot{z}_u & = A_{z} ^ \top z_u + c_0 ^ \top u 
\end{aligned}
\end{equation}
where $A_{z} \triangleq A_0 - b_{z} c_0$ and $b_z \in \R ^ {n \times 1}$ is defined by the user such that $A_z$ is a Hurwitz matrix. 
Let us define 
\begin{equation} \label{eq:Es}
E_y \triangleq \mathcal{C}_{y} \mathcal{C}_{0} ^ {- 1}
, \quad
E_u \triangleq \mathcal{C}_{u} \mathcal{C}_{0} ^ {- 1}
\end{equation}
where $\mathcal{C}_{0}$ is the controllability matrix of $(A_{z} ^ \top, \ c_0 ^ \top)$, $\mathcal{C}_{y}$ is the controllability matrix of $(A_{z} ^ \top, \ z_y)$, and $\mathcal{C}_{u}$ is the controllability matrix of $(A_{z} ^ \top, \ z_u)$.
\begin{lemma}
$z_y$ and $z_u$ are the states of the identifier-based estimator in \eqref{eq:model-0}. $E_y$ and $E_u$ are defined as \eqref{eq:Es}. If we define
a state estimate $\hat{x}_p$ for $\Sigma_p$ in \eqref{eq:model-siso} as
\begin{equation} \label{eq:xs}
\hat{x}_{p} = E_y ^ \top (b_y + b_{z}) + E_u ^ \top b_u,
\end{equation}
the state estimation error $e \triangleq x_p - \hat{x}_p$ follows 
\begin{equation} \label{eq:model-error}
\dot{e} = A_{z} e - b_{z} w.
\end{equation}
\end{lemma}
\begin{proof}
This is similar to the proofs of \cite[Lemma 2]{morse1980global} and \cite[Lemma 1]{he2020robust}. 
Subtracting \eqref{eq:model-error} from \eqref{eq:model-siso}, we obtain that
\begin{equation} \label{eq:model-siso-hat}
\dot{\hat{x}}_p = A_{z} \hat{x}_p + (b_y + b_{z}) y + b_u u.
\end{equation}
Since $\mathcal{C}_{0}$ is the controllability matrix of $(A_z ^ \top, \, c_0 ^ \top)$, we can derive from \eqref{eq:model-0} that 
\begin{equation} \label{L-1-1}
\begin{aligned} 
\dot{E}_{y} & = A_z ^ \top E_{y} + y \cdot \I, \\
\dot{E}_{u} & = A_z ^ \top E_{u} + u \cdot \I.
\end{aligned}
\end{equation}
Notice that $A_z$ is in a canonical form, which leads to $E_y ^ \top A_z = A_z E_y ^ \top$ and $E_u ^ \top A_z = A_z E_u ^ \top$. By taking the transpose of \eqref{L-1-1}, we obtain $\dot{E}_{y} ^ \top = A_z E_{y} ^ \top + y \cdot \I$ and $\dot{E}_{u} ^ \top = A_z E_{u} ^ \top + u \cdot \I$.
Therefore, \eqref{eq:model-siso-hat} holds if $\hat{x}_p = E_y ^ \top (b_y + b_{z}) + E_u ^ \top b_u$.
\end{proof}

Notice that in \eqref{eq:xs}, $\hat{x}_p$ is affine in $b_y$ and $b_u$, which are defined in \eqref{eq:c-limit}.
Substituting \eqref{eq:c-limit} into \eqref{eq:xs}, we have
\begin{equation} \label{eq:xs-bar}
\begin{aligned}
& \hat{x}_p = 
\bar{x}_p + \sum_{i} ^ {n_p} \delta_i \cdot \tilde{x}_i ^ {p} 
\\
& \bar{x}_p \triangleq 
E_y ^ \top ( \bar{b}_y + b_{z} ) + E_u ^ \top \bar{b}_u, 
\quad
  \tilde{x}_i ^ {p}  \triangleq 
\theta_i ^ y \cdot E_y ^ \top \tilde{b}_i  + \theta_i ^ u \cdot E_u ^ \top \tilde{b}_i
\end{aligned}
\end{equation}
where $\delta \triangleq [\delta_1 \ \delta_2 \ \cdots \ \delta_{n_p}]$ and $\abs{\delta_i} \leq 1$ for all $i = 1, \, 2, \, \cdots, \, n_p$.
Let us define a set
\begin{equation} \label{eq:delta}
\begin{aligned}
\bar{\Delta} \triangleq \big\{ [\delta_1 \ \delta_2 \ \cdots \ \delta_{n_p}] : 
\abs{\delta_i} = 1, \, i = 1, \, 2, \, \cdots, \, n_p \big\},
\end{aligned}
\end{equation}
which consists of all the $2 ^ {n_p}$ extreme values of $\delta$.
Then, 
we have
\begin{equation} \label{eq:xp-polytopic}
\hat{x}_p ( \delta ) \in \mathsf{Co} \big\{ \bar{x}_p + \sum_{i} ^ {n_p} \delta_i \cdot \tilde{x}_i ^ {p}, \, \forall \, \delta \in \bar{\Delta} \big\}.
\end{equation}
Although we do not know the exact value of $\hat{x}_p$ due to the uncertainty in $b_y$ and $b_u$, 
\eqref{eq:xp-polytopic} shows that the possible value of $\hat{x}_p (\delta)$ is in a polytopic region in $\R ^ n$. 

\subsection{Barrier Function Estimation}

Next, we will explain how \eqref{eq:model-error} and \eqref{eq:xp-polytopic} help us find a computable upper bound for our barrier function $\mathbf{B}$.

\begin{proposition} \label{prop:b-bar}
Let us define
\begin{equation} \label{eq:V-bar}
\bar{\mathbf{B}} \triangleq 
\underset{\bar{\delta} \in \bar{\Delta}}{\mathsf{max}} \ 
\mathbf{B} ( \hat{x}_\mathsf{CL} (\bar{\delta}) )
+ r_{e},
\end{equation}
where $\hat{x}_\mathsf{CL} (\bar{\delta}) \triangleq [ \hat{x}_{p} (\bar{\delta}) ^ \top \ x_{k} ^ \top ] ^ \top$ and $r_{e} > 0$.
Supposing that $\abs{w} \leq \bar{w}$, $\mathbf{B} (x_\mathsf{CL}) \leq \bar{\mathbf{B}}$ if there exists $\mu_e \geq 0$ such that
\begin{equation} \label{eq:lmi-estimator}
\begin{bmatrix}
A_{z} ^ \top X + X A_{z} + \mu_e \cdot X & \star \\
- b_{z} ^ \top X & - \mu_e \cdot \frac{r_{e} ^ 2}{\bar{w} ^ 2}
\end{bmatrix}
\prec \0.
\end{equation}
\end{proposition}
\begin{proof}
Through the triangle inequality of $\norm{\star}_P$, we have
\begin{equation}
\mathbf{B} (x_\mathsf{CL})
=
\norm{x_\mathsf{CL}}_{P}
\leq 
\norm{\hat{x}_\mathsf{CL} (\delta)}_{P} + \norm{S_p ^ \top e}_{P},
\end{equation}
where $e \triangleq x_{p} - \hat{x}_{p}$.
According to \eqref{eq:xp-polytopic},
\begin{equation}
\norm{\hat{x}_\mathsf{CL} (\delta)}_{P} \leq \underset{\bar{\delta} \in \bar{\Delta}}{\mathsf{max}} \ \norm{\hat{x}_\mathsf{CL} (\bar{\delta})}_{P}.
\end{equation}
Based on the definition of $P$ in \eqref{eq:P},
\begin{equation}
\norm{S_p ^ \top e}_{P}
\leq
\norm{e}_{X}.
\end{equation}
Therefore, $\mathbf{B} (x_\mathsf{CL}) \leq \bar{\mathbf{B}}$ for all $w$ that $\abs{w} \leq \bar{w}$ if $\underset{\bar{\delta} \in \bar{\Delta}}{\mathsf{max}} \ \norm{\hat{x}_\mathsf{CL} (\bar{\delta})}_c + \norm{e}_{X} \leq \bar{\mathbf{B}}$, or equivalently $\norm{e}_{X} \leq r_{e}$, for all $w$ that $\abs{w} \leq \bar{w}$.

At $t = 0$, $e \triangleq x_p - \hat{x}_p = 0$ for $x_p = z_y = z_u = \0$.
Since $e = 0$  at $t = 0$ and $\dot{e} = A_{z} e - b_{z} w$, $\norm{e}_{X} \leq r_{e}$ for all $w$ that $\abs{w} \leq \bar{w}$ if and only if $\frac{\mathsf{d} \, \norm{e}_{X} ^ 2}{\mathsf{d} \, t} < 0$, or equivalently
\begin{equation} \label{eq:B-dot-1}
\begin{bmatrix}
e \\
w 
\end{bmatrix}
^ \top 
\begin{bmatrix*}
A_z ^ \top X + X A_z & \star \\
- b_z ^ \top X & \0
\end{bmatrix*}
\begin{bmatrix}
e \\
w 
\end{bmatrix}
< 0
,
\end{equation}
for all $e$ and $w$ that
\begin{equation} \label{eq:s-condition-2}
e ^ \top X e \geq r_e ^ 2
\quad \text{and} \quad
w ^ 2 \leq \bar{w} ^ 2.
\end{equation}
Using the S-procedure, we obtain that \eqref{eq:B-dot-1} holds under the conditions in \eqref{eq:s-condition-2}
if there exists $\mu_e \geq 0$ such that \eqref{eq:lmi-estimator} holds.
Therefore, $\mathbf{B} (x_\mathsf{CL}) \leq \bar{\mathbf{B}}$ for all $w$ that $\abs{w} \leq \bar{w}$ if there exists $\mu_e \geq 0$ such that \eqref{eq:lmi-estimator} holds.
\end{proof}

Notice that $X$ is already obtained from optimization \eqref{eq:optimization-volume}. If we fix the value of $\mu_e$, \eqref{eq:lmi-estimator} becomes an LMI in $b_z$.
Through convex optimization
\begin{equation} \label{eq:optimization-estimator}
\begin{aligned}
& \underset{b_{z}}{\mathsf{minimize}}
& & r_{e} ^ 2 \\
& \mathsf{subject \ to} & & 
\eqref{eq:lmi-estimator}
\end{aligned}
\end{equation}
$b_{z}$ in \eqref{eq:model-0} can be defined for minimizing $r_{e}$. 
Since $\hat{x}_\mathsf{CL} (\bar{\delta})$ for all $\bar{\delta} \in \bar{\Delta}$ can be obtained from our identifier-based estimator, $\bar{\mathbf{B}}$ is available to us.

\subsection{Switching Logic of $\Sigma_s$} \label{sec:switching}

\begin{figure}[!tbp]
\centering
\def\svgwidth{0.49\textwidth}
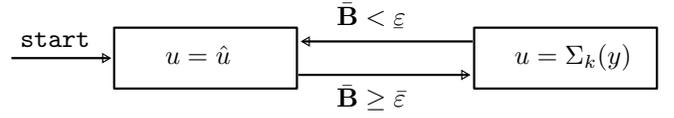
\caption{$\Sigma_s$ switches from the original input $u = \hat{u}$ to $u = \Sigma_k (y)$ if $\bar{\mathbf{B}} \geq \bar{\varepsilon}$ and switches back to $u = \hat{u}$ if $\bar{\mathbf{B}} < \ubar{\varepsilon}$.}
\vspace{-5pt}
\label{fig:switching}
\end{figure}

Based on the value of $\bar{\mathbf{B}}$ (Fig.~\ref{fig:switching}), we can now define the switching logic of $\Sigma_s$.
Let us define two thresholds $\ubar{\varepsilon}$ and $\bar{\varepsilon}$ (with $\varepsilon < \ubar{\varepsilon} < \bar{\varepsilon} \leq 1$).
$\Sigma_s$ switches from the original input $u = \hat{u}$ to $u = \Sigma_k (y)$ if $\bar{\mathbf{B}} \geq \bar{\varepsilon}$ and switches back to $u = \hat{u}$ if $\bar{\mathbf{B}} < \ubar{\varepsilon}$.
According to Proposition~\ref{prop:CL-convex}, $x_\mathsf{CL}$ converges to residual set $\{ x_\mathsf{CL} : \mathbf{B} (x_\mathsf{CL}) \leq \varepsilon \}$ when $u = \Sigma_k (y)$. Therefore, when $\Sigma_k$ is in control of the system, the true value of $\mathbf{B} (x_\mathsf{CL})$ goes below $\ubar{\varepsilon}$ in finite time. By setting the values of $\ubar{\varepsilon}$ and $\bar{\varepsilon}$ closer to 1, we reduce the intervention from $\Sigma_k$ in the original operation of $\Sigma_p$.

\begin{figure}[!tbp]
\centering
\includestandalone[width=.22\textwidth]{fig-5}
\caption{In our example, we consider a simplified $1$-DOF physical human-robot interaction system, which a mass-spring system with an uncertain human stiffness $k$ and a robot inertia $m$.}
\vspace{-5pt}
\label{fig:mass-spring}
\end{figure}

\begin{figure*}
\footnotesize
\centering
\def\svgwidth{1.0\textwidth}
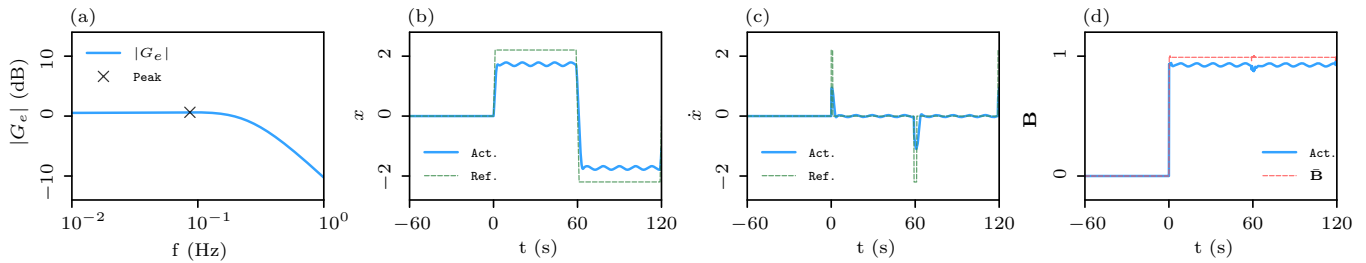
\vspace{-15pt}
\caption{{\bf Simulation Results of Example}---(a) shows the frequency response of $G_e (s)$, where the peak value is at $f_e = 0.09 \, \mathrm{Hz}$. 
(b)-(d) show the results of our safety control test under the disturbance signals $w (t) = 0.05 \sin ( 2 \pi f_e t )$.
In particular,
(b)-(c) show the actual ($\mathtt{Act.}$) and reference ($\mathtt{Ref.}$) values of $x$ and $\dot{x}$ and (d) shows the actual value ($\mathtt{Act.}$) and upper bound ($\bar{\mathbf{B}}$) of $\mathbf{B}$.}
\vspace{-5pt}
\label{fig:example}
\end{figure*}

\section{Example} \label{sec:example}

In this section, we provide an example to illustrate the robust fault detection and safety control of our proposed method. 

\subsection{System Model}
Here we consider an uncertain mass-spring system $\Sigma_p$ (Fig.~\ref{fig:mass-spring}) with a transfer function 
\begin{align}
G_p (s) = \frac{y (s)}{u (s)} = \frac{k}{m \cdot s ^ 2 + k}
\end{align}
where the system has a unit mass $m = 1$ and a spring stiffness $k$, the output $y = k (x - x_0)$ measures the spring force, and the input $u$ represents an adjustable force exerting to the mass. 
The spring stiffness is uncertain and defined as
$
k = \hat{k} + \delta \cdot \bar{k}
$,
where $\hat{k} = 10$, $\bar{k} = 1$, and $\abs{\delta} \leq 1$. 
The state model of $\Sigma_p$ can be expressed as
\begin{align}
\dot{x}_p
& = 
\begin{bmatrix}
0 & 0 \\
1 & 0
\end{bmatrix}
x_p
+
\begin{bmatrix}
- \frac{k}{m} \\
0
\end{bmatrix}
y 
+ 
\begin{bmatrix}
\frac{k}{m} \\
0
\end{bmatrix}
u 
\label{eq:model-example}
\\
y & = 
\begin{bmatrix}
0 & 1
\end{bmatrix}
x_p
+ w
\notag
\end{align}
where $x_p = [ \dot{x} \ x ] ^ \top$ and $w$ is an unknown exogenous
input.
If we consider this system as a simplified $1$-DOF human-robot interaction model in a wearable robot control problem \cite{thomas2021formulating}, then $m$ is the robot inertia, $k$ is an uncertain human joint stiffness, and $x_0 = \frac{w}{k}$ is a desired joint position where the human operator tends to move to. 

The safe regions $\mathcal{X}_s$ and $\mathcal{U}$ are defined as
\begin{alignat}{4}
\mathcal{X}_s & \triangleq \{ [\dot{x} \ x] ^ \top \ & : \ && |\dot{x}| & \leq 2, \ |x| & \leq 2 \}, \label{eq:x-limit-example} \\
\mathcal{U}   & \triangleq \{  u \ & : \ &&           |u| & \leq 10 \}. 
\end{alignat}
Assuming that $\abs{w} \leq \bar{w} = 0.05$, we aim to achieve a residual set $\{ x_\mathsf{CL} : \mathbf{B} (x_\mathsf{CL}) \leq \varepsilon \}$ of our barrier function with $\varepsilon = 0.5$. 

\subsection{Safety Control} \label{sec:safety-control}

Through optimizations \eqref{eq:optimization-volume} and \eqref{eq:optimization-estimator}, we obtain a barrier pair $(\mathbf{B}, \, \Sigma_k)$ and a state estimator in the form of \eqref{eq:model-0}, which provides us with a state estimate $\hat{x}_p$ to calculate a barrier function upper bound $\bar{\mathbf{B}}$ with $r_{e} = 0.05$. 

In our safety control test, we want to demonstrate the robustness of our switching system $\Sigma_s$  with respect to the bounded model uncertainty and disturbance in the test.
Unfortunately, it is difficult to implement the exact worst case of the bounded disturbance signal $w (t)$ that maximizes the peak value of the estimation error residue $\norm{e}_{X}$. Instead, we implement $w (t)$ as a sinusoidal signal in this example.
Let us define a transfer function
\begin{equation} \label{eq:Ge}
G_{e} (s) \triangleq \frac{\norm{e (s)}_{X}}{w (s)}
\end{equation}
from the disturbance $w (s)$ to the estimation error residue $\norm{e (s)}_{X}$.
In the frequency domain (Fig.~\ref{fig:example}a), $\norm{G_{e} (s)}_{\infty} = 1.09$ at $f_e = 0.09 \, \mathrm{Hz}$. Therefore, we define the disturbance as a sinusoidal signal $w (t) = \bar{w} \cdot \sin ( 2 \pi f_e t )$, which gives us the maximum sinusoidal response of $G_{e} (s)$.

To test the fault detection, we implement the original input $\hat{u}$ as a reference tracking controller, which lets $x$ follow a trapezoidal reference (Fig.~\ref{fig:example}b-c). The reference is designed to violate the safety limits of $x$ and $\dot{x}$ on purpose in such a way that $\hat{u}$ can cause potential risks.
Fig.~\ref{fig:example}b-d show that the safety controller $\Sigma_k$ is correctly triggered when the system is about to violate the constraints of $x$ and $\dot{x}$. Fig.~\ref{fig:example}d shows that the true value of $\mathbf{B}$ is strictly lower than $\bar{\mathbf{B}}$ at all times.

\section{Discussion} \label{sec:discussion}

In our problem statement, we assume that the initial state of $x_p$ is $\0$. Similar to Ref.~\cite{he2020robust}, our fault detector can also be extended to cases where the initial states of $x_p$ are not $\0$. The Hurwitz matrix $A_z$ in our identifier-based estimator in \eqref{eq:model-0} guarantees that the state estimation error due to an unknown initial state of $x_p$ converges to $\0$ exponentially.

Fig.~\ref{fig:example}d in our example shows a slight over-conservatism of the barrier function upper bound $\bar{\mathbf{B}}$. This is partly because the worst case of the disturbance input $w (t)$ is difficult to find.
In our example, we implement the disturbance input $w (t)$ as a sinusoidal signal, which only leads to the worst case of the sinusoidal response of $G_{e} (s)$ in \eqref{eq:Ge}.
Note that as long as $w \leq \bar{w}$, constraint \eqref{eq:lmi-estimator} guarantees $\norm{e (t)}_{X} \leq r_e$ no matter what type of signal we implement $w (t)$ as. 

In this paper, we considered the safety control problem for an uncertain SISO system with partial state information. 
By knowing the limits of the uncertain model parameters and disturbance a priori, our fault detector and safety controller work together to protect the uncertain SISO system from potential risks.
In the future, we will extend our safety control method to MIMO systems.

\bibliographystyle{IEEEtran}
\bibliography{main}

\end{document}

%% file: fig-3.pdf_tex
%% Creator: Inkscape 1.1 (c4e8f9e, 2021-05-24), www.inkscape.org
%% Accompanies image file 'fig-3.pdf' (pdf, eps, ps)
\begingroup%
  \makeatletter%
  \providecommand\color[2][]{%
    \errmessage{(Inkscape) Color is used for the text in Inkscape, but the package 'color.sty' is not loaded}%
    \renewcommand\color[2][]{}%
  }%
  \providecommand\transparent[1]{%
    \errmessage{(Inkscape) Transparency is used (non-zero) for the text in Inkscape, but the package 'transparent.sty' is not loaded}%
    \renewcommand\transparent[1]{}%
  }%
  \providecommand\rotatebox[2]{#2}%
  \newcommand*\fsize{\dimexpr\f@size pt\relax}%
  \newcommand*\lineheight[1]{\fontsize{\fsize}{#1\fsize}\selectfont}%
  \ifx\svgwidth\undefined%
    \setlength{\unitlength}{180.88070297bp}%
    \ifx\svgscale\undefined%
      \relax%
    \else%
      \setlength{\unitlength}{\unitlength * \real{\svgscale}}%
    \fi%
  \else%
    \setlength{\unitlength}{\svgwidth}%
  \fi%
  \global\let\svgwidth\undefined%
  \global\let\svgscale\undefined%
  \makeatother%
  \begin{picture}(1,0.23013178)%
    \lineheight{1}%
    \setlength\tabcolsep{0pt}%
    \put(0,0){\includegraphics[width=\unitlength,page=1]{fig-3.pdf}}%
    \put(0.92194708,0.16441183){\makebox(0,0)[lt]{\lineheight{1.25}\smash{$w$}}}%
    \put(-0.00076186,0.16491822){\makebox(0,0)[lt]{\lineheight{1.25}\smash{$\hat{u}$}}}%
    \put(0.48210799,0.16491822){\makebox(0,0)[lt]{\lineheight{1.25}\smash{$u$}}}%
    \put(0.48292862,0.05779919){\makebox(0,0)[lt]{\lineheight{1.25}\smash{$y$}}}%
    \put(0.68635516,0.13298298){\makebox(0,0)[lt]{\lineheight{1.25}\smash{$\Sigma_p$}}}%
    \put(0.20469932,0.16622061){\makebox(0,0)[lt]{\lineheight{1.25}\smash{$u = \hat{u}$}}}%
    \put(0.24214645,0.11085592){\makebox(0,0)[lt]{\lineheight{1.25}\smash{$\mathrm{or}$}}}%
    \put(0.20469932,0.04882638){\makebox(0,0)[lt]{\lineheight{1.25}\smash{$u = \Sigma_k (y)$}}}%
    \put(0.05642827,0.19262754){\makebox(0,0)[lt]{\lineheight{1.25}\smash{$\Sigma_s$}}}%
  \end{picture}%
\endgroup%

%% file: fig-4.pdf_tex
%% Creator: Inkscape 1.1 (c4e8f9e, 2021-05-24), www.inkscape.org
%% Accompanies image file 'fig-4.pdf' (pdf, eps, ps)
\begingroup%
  \makeatletter%
  \providecommand\color[2][]{%
    \errmessage{(Inkscape) Color is used for the text in Inkscape, but the package 'color.sty' is not loaded}%
    \renewcommand\color[2][]{}%
  }%
  \providecommand\transparent[1]{%
    \errmessage{(Inkscape) Transparency is used (non-zero) for the text in Inkscape, but the package 'transparent.sty' is not loaded}%
    \renewcommand\transparent[1]{}%
  }%
  \providecommand\rotatebox[2]{#2}%
  \newcommand*\fsize{\dimexpr\f@size pt\relax}%
  \newcommand*\lineheight[1]{\fontsize{\fsize}{#1\fsize}\selectfont}%
  \ifx\svgwidth\undefined%
    \setlength{\unitlength}{180.88070297bp}%
    \ifx\svgscale\undefined%
      \relax%
    \else%
      \setlength{\unitlength}{\unitlength * \real{\svgscale}}%
    \fi%
  \else%
    \setlength{\unitlength}{\svgwidth}%
  \fi%
  \global\let\svgwidth\undefined%
  \global\let\svgscale\undefined%
  \makeatother%
  \begin{picture}(1,0.16793611)%
    \lineheight{1}%
    \setlength\tabcolsep{0pt}%
    \put(0,0){\includegraphics[width=\unitlength,page=1]{fig-4.pdf}}%
    \put(0.50132717,0.00935924){\makebox(0,0)[lt]{\lineheight{1.25}\smash{$\bar{\mathbf{B}} \geq \bar{\varepsilon}$}}}%
    \put(0.23971351,0.0751676){\makebox(0,0)[lt]{\lineheight{1.25}\smash{$u = \hat{u}$}}}%
    \put(0.77296137,0.0751676){\makebox(0,0)[lt]{\lineheight{1.25}\smash{$u = \Sigma_k (y)$}}}%
    \put(0.50132717,0.13467815){\makebox(0,0)[lt]{\lineheight{1.25}\smash{$\bar{\mathbf{B}} < \ubar{\varepsilon}$}}}%
    \put(0.01879404,0.10100411){\makebox(0,0)[lt]{\lineheight{1.25}\smash{$\mathtt{start}$}}}%
  \end{picture}%
\endgroup%

%% file: fig-5.tex
\begin{tikzpicture}
\tikzstyle{spring}=[thick,decorate,decoration={zigzag,pre length=0.2cm,post length=0.2cm,segment length=6}]
\tikzstyle{damper}=[thick,decoration={markings,  
  mark connection node=dmp,
  mark=at position 0.5 with 
  {
    \node (dmp) [thick,inner sep=0pt,transform shape,rotate=-90,minimum width= 5pt,minimum height=1pt,draw=none] {};
    \draw [thick] ($(dmp.north east)+(2pt,0)$) -- (dmp.south east) -- (dmp.south west) -- ($(dmp.north west)+(2pt,0)$);
    \draw [thick] ($(dmp.north)+(0,-2pt)$) -- ($(dmp.north)+(0,2pt)$);
  }
}, decorate]
\tikzstyle{ground}=[fill, pattern=north east lines, draw=none,minimum width=0.75cm, minimum height=0.15cm]

\begin{scope}[xshift=5cm]
\node [draw,outer sep=0pt,thick] (M1) [minimum width=1.0cm, minimum height=0.6cm] {\footnotesize $m$};

\node (ground1) [ground,anchor=north, yshift=-0.25cm, minimum width=1.5cm] at (M1.south) {};

\draw (ground1.north east) -- (ground1.north west);

\draw (M1.south west) ++ (0.2cm,-0.125cm) circle (0.125cm)  (M1.south east) ++ (-0.2cm,-0.125cm) circle (0.125cm);

\node (wall1) [ground, rotate=-90, minimum width = 0.6cm, yshift = - 1.6cm] {};

\draw (wall1.north east) -- (wall1.north west);

% \node (wall2) [ground, rotate=-90, minimum width = 0.56cm, yshift =   1.6cm] {};

% \draw (wall2.south east) -- (wall2.south west);

% \draw [damper] ($(M1.north east)!(wall2.90)!(M1.south east)$) -- ($(wall2.south east)!(wall2.90)!(wall2.south west)$) node [midway,below] {\footnotesize $b_e$};

\draw [spring] ($(wall1.north east)!(wall1.90)!(wall1.north west)$) -- ($(M1.north west)!(wall1.90)!(M1.south west)$) node [midway,below] {\footnotesize $k$};

\draw [-latex,ultra thick] ($(M1.north east)$) ++ (0.cm,-0.0cm) -- +( .5cm,0cm) node [midway,above] {\footnotesize $u$};
% \draw [-latex,ultra thick] ($(M1.north west)$) ++ (-.6cm,-.0cm) -- +( .6cm,0cm);
% \draw [-latex,ultra thick] ($(M1.north west)$) ++ (-.6cm,-.0cm) -- +(-.6cm,0cm) node [pos=.0,above] { $y$};

 \draw[thick, dashed] ($(M1.north west)$) -- ($(M1.north west) + (0,.5)$);
 \draw[thick, -latex] ($(M1.north west) + (0,0.25)$) -- ($(M1.north west) + (0.5,0.25)$) node [midway, above] {\footnotesize $x$};
  \draw[thick, dashed] ($(wall1.north west)$) -- ($(wall1.north west) + (0,.5)$);
 \draw[thick, -latex] ($(wall1.north west) + (0,0.25)$) -- ($(wall1.north west) + (0.5,0.25)$) node [midway, above] {\footnotesize $x_0$};

\end{scope}
\end{tikzpicture}

%% file: fig-8.pdf_tex
%% Creator: Inkscape 1.1 (c4e8f9e, 2021-05-24), www.inkscape.org
%% Accompanies image file 'fig-8.pdf' (pdf, eps, ps)
\begingroup%
  \makeatletter%
  \providecommand\color[2][]{%
    \errmessage{(Inkscape) Color is used for the text in Inkscape, but the package 'color.sty' is not loaded}%
    \renewcommand\color[2][]{}%
  }%
  \providecommand\transparent[1]{%
    \errmessage{(Inkscape) Transparency is used (non-zero) for the text in Inkscape, but the package 'transparent.sty' is not loaded}%
    \renewcommand\transparent[1]{}%
  }%
  \providecommand\rotatebox[2]{#2}%
  \newcommand*\fsize{\dimexpr\f@size pt\relax}%
  \newcommand*\lineheight[1]{\fontsize{\fsize}{#1\fsize}\selectfont}%
  \ifx\svgwidth\undefined%
    \setlength{\unitlength}{643.99273682bp}%
    \ifx\svgscale\undefined%
      \relax%
    \else%
      \setlength{\unitlength}{\unitlength * \real{\svgscale}}%
    \fi%
  \else%
    \setlength{\unitlength}{\svgwidth}%
  \fi%
  \global\let\svgwidth\undefined%
  \global\let\svgscale\undefined%
  \makeatother%
  \begin{picture}(1,0.18739299)%
    \lineheight{1}%
    \setlength\tabcolsep{0pt}%
    \put(0,0){\includegraphics[width=\unitlength,page=1]{fig-8.pdf}}%
    \put(0.036297,0.02296464){\color[rgb]{0,0,0}\makebox(0,0)[lt]{\lineheight{1.25}\smash{\scriptsize $10 ^ {-2}$}}}%
    \put(0.13047388,0.02296464){\color[rgb]{0,0,0}\makebox(0,0)[lt]{\lineheight{1.25}\smash{\scriptsize $10 ^ {-1}$}}}%
    \put(0.22903017,0.02296464){\color[rgb]{0,0,0}\makebox(0,0)[lt]{\lineheight{1.25}\smash{\scriptsize $10 ^ { 0}$}}}%
    \put(0.11932517,0.00265676){\color[rgb]{0,0,0}\makebox(0,0)[lt]{\lineheight{1.25}\smash{\scriptsize $\mathrm{f \ (Hz)}$}}}%
    \put(0.03036477,0.04364478){\color[rgb]{0,0,0}\rotatebox{90}{\makebox(0,0)[lt]{\lineheight{1.25}\smash{\scriptsize $-10$}}}}%
    \put(0.03036477,0.10501381){\color[rgb]{0,0,0}\rotatebox{90}{\makebox(0,0)[lt]{\lineheight{1.25}\smash{\scriptsize $0$}}}}%
    \put(0.03036477,0.14209588){\color[rgb]{0,0,0}\rotatebox{90}{\makebox(0,0)[lt]{\lineheight{1.25}\smash{\scriptsize $10$}}}}%
    \put(0.00862539,0.08487577){\color[rgb]{0,0,0}\rotatebox{90}{\makebox(0,0)[lt]{\lineheight{1.25}\smash{\scriptsize $\abs{G_e} \ \mathrm{(dB)}$}}}}%
    \put(0.09052169,0.15139195){\color[rgb]{0,0,0}\makebox(0,0)[lt]{\lineheight{1.25}\smash{\tiny $\abs{G_e}$}}}%
    \put(0.09052169,0.13381606){\color[rgb]{0,0,0}\makebox(0,0)[lt]{\lineheight{1.25}\smash{\tiny $\mathtt{Peak}$}}}%
    \put(0.28489407,0.02296464){\color[rgb]{0,0,0}\makebox(0,0)[lt]{\lineheight{1.25}\smash{\scriptsize $-60$}}}%
    \put(0.3559401,0.02296464){\color[rgb]{0,0,0}\makebox(0,0)[lt]{\lineheight{1.25}\smash{\scriptsize $0$}}}%
    \put(0.41484267,0.02296464){\color[rgb]{0,0,0}\makebox(0,0)[lt]{\lineheight{1.25}\smash{\scriptsize $60$}}}%
    \put(0.4737452,0.02296464){\color[rgb]{0,0,0}\makebox(0,0)[lt]{\lineheight{1.25}\smash{\scriptsize $120$}}}%
    \put(0.37879193,0.00498598){\color[rgb]{0,0,0}\makebox(0,0)[lt]{\lineheight{1.25}\smash{\scriptsize $\mathrm{t \ (s)}$}}}%
    \put(0.28284387,0.05140883){\color[rgb]{0,0,0}\rotatebox{90}{\makebox(0,0)[lt]{\lineheight{1.25}\smash{\scriptsize $-2$}}}}%
    \put(0.28284387,0.10501381){\color[rgb]{0,0,0}\rotatebox{90}{\makebox(0,0)[lt]{\lineheight{1.25}\smash{\scriptsize $0$}}}}%
    \put(0.28284387,0.14985994){\color[rgb]{0,0,0}\rotatebox{90}{\makebox(0,0)[lt]{\lineheight{1.25}\smash{\scriptsize $2$}}}}%
    \put(0.26243894,0.10350953){\color[rgb]{0,0,0}\rotatebox{90}{\makebox(0,0)[lt]{\lineheight{1.25}\smash{\scriptsize $x$}}}}%
    \put(0.3430008,0.07712383){\color[rgb]{0,0,0}\makebox(0,0)[lt]{\lineheight{1.25}\smash{\tiny $\mathtt{Act.}$}}}%
    \put(0.3430008,0.05954793){\color[rgb]{0,0,0}\makebox(0,0)[lt]{\lineheight{1.25}\smash{\tiny $\mathtt{Ref.}$}}}%
    \put(0.53737317,0.02296464){\color[rgb]{0,0,0}\makebox(0,0)[lt]{\lineheight{1.25}\smash{\scriptsize $-60$}}}%
    \put(0.60841918,0.02296464){\color[rgb]{0,0,0}\makebox(0,0)[lt]{\lineheight{1.25}\smash{\scriptsize $0$}}}%
    \put(0.66732175,0.02296464){\color[rgb]{0,0,0}\makebox(0,0)[lt]{\lineheight{1.25}\smash{\scriptsize $60$}}}%
    \put(0.72622432,0.02296464){\color[rgb]{0,0,0}\makebox(0,0)[lt]{\lineheight{1.25}\smash{\scriptsize $120$}}}%
    \put(0.63127101,0.00498598){\color[rgb]{0,0,0}\makebox(0,0)[lt]{\lineheight{1.25}\smash{\scriptsize $\mathrm{t \ (s)}$}}}%
    \put(0.53532297,0.05140883){\color[rgb]{0,0,0}\rotatebox{90}{\makebox(0,0)[lt]{\lineheight{1.25}\smash{\scriptsize $-2$}}}}%
    \put(0.53532297,0.10501381){\color[rgb]{0,0,0}\rotatebox{90}{\makebox(0,0)[lt]{\lineheight{1.25}\smash{\scriptsize $0$}}}}%
    \put(0.53532297,0.14985994){\color[rgb]{0,0,0}\rotatebox{90}{\makebox(0,0)[lt]{\lineheight{1.25}\smash{\scriptsize $2$}}}}%
    \put(0.51491803,0.10350953){\color[rgb]{0,0,0}\rotatebox{90}{\makebox(0,0)[lt]{\lineheight{1.25}\smash{\scriptsize $\dot{x}$}}}}%
    \put(0.5954799,0.07712383){\color[rgb]{0,0,0}\makebox(0,0)[lt]{\lineheight{1.25}\smash{\tiny $\mathtt{Act.}$}}}%
    \put(0.5954799,0.05954793){\color[rgb]{0,0,0}\makebox(0,0)[lt]{\lineheight{1.25}\smash{\tiny $\mathtt{Ref.}$}}}%
    \put(0.78985225,0.02296464){\color[rgb]{0,0,0}\makebox(0,0)[lt]{\lineheight{1.25}\smash{\scriptsize $-60$}}}%
    \put(0.8608983,0.02296464){\color[rgb]{0,0,0}\makebox(0,0)[lt]{\lineheight{1.25}\smash{\scriptsize $0$}}}%
    \put(0.91980083,0.02296464){\color[rgb]{0,0,0}\makebox(0,0)[lt]{\lineheight{1.25}\smash{\scriptsize $60$}}}%
    \put(0.97870345,0.02296464){\color[rgb]{0,0,0}\makebox(0,0)[lt]{\lineheight{1.25}\smash{\scriptsize $120$}}}%
    \put(0.88375011,0.00498598){\color[rgb]{0,0,0}\makebox(0,0)[lt]{\lineheight{1.25}\smash{\scriptsize $\mathrm{t \ (s)}$}}}%
    \put(0.78780205,0.06016767){\color[rgb]{0,0,0}\rotatebox{90}{\makebox(0,0)[lt]{\lineheight{1.25}\smash{\scriptsize $0$}}}}%
    \put(0.78780205,0.14985994){\color[rgb]{0,0,0}\rotatebox{90}{\makebox(0,0)[lt]{\lineheight{1.25}\smash{\scriptsize $1$}}}}%
    \put(0.76557742,0.09885111){\color[rgb]{0,0,0}\rotatebox{90}{\makebox(0,0)[lt]{\lineheight{1.25}\smash{\scriptsize $\mathbf{B}$}}}}%
    \put(0.96985237,0.07712383){\color[rgb]{0,0,0}\makebox(0,0)[lt]{\lineheight{1.25}\smash{\tiny $\mathtt{Act.}$}}}%
    \put(0.96985237,0.05954793){\color[rgb]{0,0,0}\makebox(0,0)[lt]{\lineheight{1.25}\smash{\tiny $\bar{\mathbf{B}}$}}}%
    \put(0.04283943,0.17795477){\makebox(0,0)[lt]{\lineheight{1.25}\smash{\scriptsize $\mathrm{(a)}$}}}%
    \put(0.29531852,0.17795477){\makebox(0,0)[lt]{\lineheight{1.25}\smash{\scriptsize $\mathrm{(b)}$}}}%
    \put(0.5477976,0.17795477){\makebox(0,0)[lt]{\lineheight{1.25}\smash{\scriptsize $\mathrm{(c)}$}}}%
    \put(0.80027672,0.17795477){\makebox(0,0)[lt]{\lineheight{1.25}\smash{\scriptsize $\mathrm{(d)}$}}}%
  \end{picture}%
\endgroup%